\documentclass[
    ,final            
  ]
  {aipproc}

\usepackage{bm}
\usepackage{amsmath}
\usepackage{amssymb}

\layoutstyle{8x11double}

\def\bu{\bm u}

\def\aj{\rm {AJ}}			
		
\def\apj{\rm{ApJ}}

\def\aap{\rm{A\&A}}

\def\mnras{\rm{MNRAS}}

\def\solphys{\rm{Solar~Phys.}}

\begin{document}

\title{Solar-type Variables}


\author{G{\"u}nter Houdek}{
  address={Institute of Astronomy, University of Vienna, A-1180 Vienna, Austria}
}



\begin{abstract}
The rich acoustic oscillation spectrum in solar-type variables make these
stars particularly interesting for studying fluid-dynamical aspects 
of the stellar interior. I present a summary of the properties of 
solar-like oscillations, how they are excited and damped and 
discuss some of the recent progress in 
using asteroseismic diagnostic techniques for analysing low-degree
acoustic modes. Also the effects of stellar-cycle variations in
low-mass main-sequence stars are addressed.
\end{abstract}

\maketitle


\section{Introduction}
\vspace{-8pt}
Solar-type stars possess extended surface convection zones. The observed
oscillation modes generally behave as acoustic modes and their frequencies
are sensitive predominantly to the sound speed $c$ in the stellar interior.
It appears that all possible oscillation modes are intrinsically stable. They
are excited stochastically by the strong emission of acoustic noise
by the turbulent velocity field in the upper convectively unstable layers of
the star. The excitation occurs in a broad frequency range, giving rise to a
rich pulsation spectrum. The amplitudes of the oscillations are small,
typically 5 ppm L$_\odot$/M$_\odot$ (Kjeldsen \& Bedding 1995). Such small
amplitudes allow us to describe the pulsations with linear theory, however,
they also pose a challenge to the detection limits of observing campaigns.
But the recently made progress in continuously lowering the detection 
threshold of today's instrumentation has allowed us to have access to
these low-amplitude pulsators, providing
high-precision data such as from the space mission CoRoT
(Convection Rotation and planetary Transits; Baglin 2003), from the
recently launched Kepler satellite (Basri et al. 2005, 
Christensen-Dalsgaard 2007), and 
from the ground-based observing campaigns, such 
as SONG (Grundahl et al. 2007) and those of the kind organized by
Kjeldsen, Bedding and their colleagues (e.g. Kjeldsen et al. 2005).
Solar-like oscillations have been convincingly observed on other stars and 
augur asteroseismic diagnosis that will raise stellar physics to a new level 
of sophistication. It will not be possible to measure either the internal 
structure or the internal motion of a distant star with a resolution 
comparable with that which we have achieved for the Sun, because high-degree 
modes will not be accessible in the foreseeable future; but some 
theoretically important properties, such as the gross structure of the 
energy-generating core and the extent to which it is convective, and the 
large-scale variation of the angular velocity, will become available. 
Such information will be of crucial importance for checking, and then 
calibrating, the theory of the structure and evolution of stars, the backbone 
of theoretical astrophysics. 

\begin{figure}[ht]
 \centering
 \includegraphics[width=0.9\linewidth]{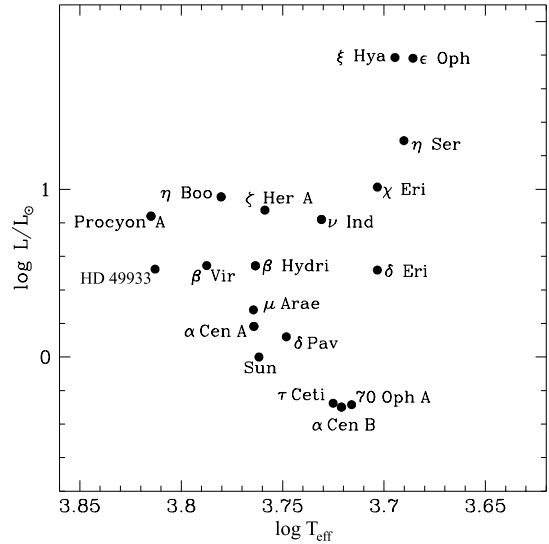}
 \caption{Hertzsprung-Russell diagram of stars in which solar-like
          oscillations have been detected (adapted from Aerts et al. 2008;
          the original figure was kindly provided by Fabien Carrier).
          }
 \label{fig:1}
\end{figure}

\begin{figure}[ht]
 \centering
 \includegraphics[width=0.9\textwidth]{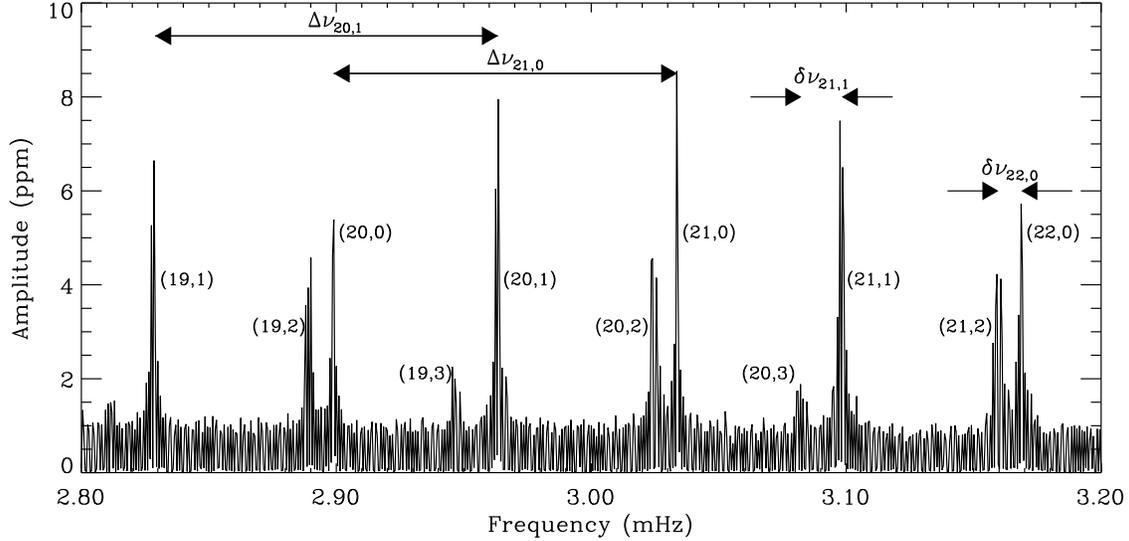}
 \caption{Small section of a solar acoustic power spectrum. 
          The radial order $n$ and spherical degree $l$ are indicated 
          in pairs of ($n$,$l$) for each mode. The large and small 
          frequency separations, $\Delta\nu_{n,l}$ and $\delta\nu_{n,l}$
          are in general functions of $n$ and $l$ and can be used to 
          infer the mass and age of a star (adapted from 
          Christensen-Dalsgaard 2001).}
 \label{fig:2}
\end{figure}

The first indication of excess acoustic power in other stars with
a frequency dependence similar to the Sun was reported 
by Brown et al. (1991) for the F5 star Procyon A ($\alpha$ CMi), in which
the first unambiguous detection of solar-like oscillations
was reported by Marti\'c et al. (1999), confirmed later
by Mosser et al. (2008). First (unconfirmed) detection of individual peaks 
in the acoustic power spectrum from high-precision time-resolved spectroscopic 
observations was published for the G5 star $\eta$ Boo by 
Kjeldsen et al. (1995), but it was not before 2003 that an unambiguous 
confirmation was established by Carrier et al. (2003) and 
Kjeldsen et al. (2003). The location of stars in the
Hertzsprung-Russell diagram, in which solar-like oscillations have been 
detected until today, are indicated in Fig.$\;$\ref{fig:1}, and a summary
of such stars was presented
recently by Bedding \& Kjeldsen (2007).

\vspace{-10pt}
\section{Solar-like oscillation properties}
\vspace{-8pt}
Only modes of low degree can be observed in distant stars, however, in
general, the observed modes are of high radial order, which allows us to
extract diagnostic properties of the frequencies $\nu_{n,l}$ 
with radial order $n$ and spherical degree $l$ for the asymptotic 
limit $n/l\rightarrow\infty$.
The diagnostic properties of this type of mode have been studied extensively
in the solar case. From asymptotic theory we find for the cyclic oscillation
frequencies (Gough 1986, 1993; see also Tassoul 1980 and Vandakurov 1967)
\begin{equation}
\nu_{n,l}\simeq\left(n+\frac{l}{2}+\hat\epsilon\right)\nu_0-\frac{AL^2-B}{\nu_{n,l}}\nu^2_0+{\rm O}(\nu^4_0)\,,
\label{eq:asymptotic}
\end{equation}
where
\begin{equation}
\nu_0=\left[2\int_0^R\frac{{\rm d}r}{c}\right]^{-1}
\end{equation}
is the inverse of twice the sound travel time between the centre and 
surface ($R$ is surface radius), and
\begin{equation}
A=\frac{1}{4\pi^2\nu_0}\left[\frac{c(R)}{R}-\int_0^R\frac{{\rm d}c}{{\rm d}r}\frac{{\rm d}r}{r}\right]\,.
\end{equation}
The frequency-dependent coefficient $\hat\epsilon$ is determined by the 
reflection properties of the surface layers, as is the small correction term
$B$, and $L^2=l(l+1)$.  The value of $\nu_0$ can be estimated from taking
the average (over $n$ and $l$) of the so-called large frequency separation
$\Delta\nu_{n,l}\equiv\nu_{n,l}-\nu_{n-1,l}$
between modes of like degree and
consecutive order. The last two terms on the right-hand side of
equation (\ref{eq:asymptotic}) lift the degeneracy between modes with
the same value of $n+l/2$ and leads to the so-called small frequency separation
$\delta\nu_{n,l}\equiv\nu_{n,l}-\nu_{n-1,l+2}$. One obtains a frequency 
structure in which modes of odd degree fall approximately halfway between 
modes of even degree, which is illustrated in Fig.\,\ref{fig:2} for a 
solar spectrum. The mean  small frequency separation (averaged over $n$)
\begin{equation}
\langle\delta\nu_{n,l}\rangle
=\langle\nu_{n,l}-\nu_{n-1,l+2}\rangle\,
\simeq\,[2A(2l+3)]\frac{\nu^2_0}{\langle\nu_{n,l}\rangle}
\label{eq:meansmallsep}
\end{equation}
is predominantly determined by the acoustic sound speed in the
stellar core and hence is sensitive to the chemical composition there and
consequently is an indicator for the stellar age (e.g., Gough 2001, 
Houdek \& Gough 2008).
\vspace{-12pt}
\subsection{Seismic diagnostic}
\begin{figure}[t]
 \centering
 \includegraphics[width=1.0\linewidth]{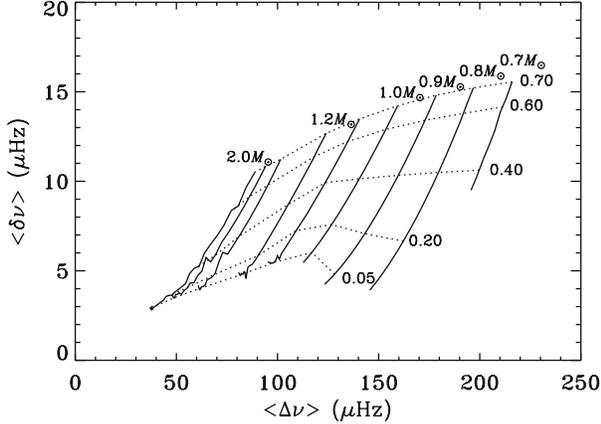}
 \caption{Stellar evolutionary tracks (solid curves, indicated by the 
          stellar mass) and curves of constant
          central hydrogen abundance (dashed curves, indicated by the central
          hydrogen abundance by mass, $X_{\rm c}$) in terms of the
          average large separation $\langle\Delta\nu\rangle$ and
          average small separation 
          $\langle\delta\nu\rangle\equiv\langle\delta\nu_{n0}\rangle$
          [c.f. equations (\ref{eq:meanlargesep}) \& (\ref{eq:meansmallsep})]
          (from Christensen-Dalsgaard 1993).}
 \label{fig:3}
\end{figure}

\vspace{-8pt}
Acoustic modes depend predominantly on the sound speed in the stellar interior
and consequently on the chemical composition. As the star evolves, the
chemical composition will vary and consequently also the value of the
small frequency separation $\delta\nu_{n,l}$. So does the mean large
frequency separation (averaged over $n$ and $l$) 
\begin{equation}
\langle\Delta\nu\rangle
=\langle\nu_{n,l}-\nu_{n-1,l}\rangle
\simeq\nu_0\propto(M/R^3)^{1/2}
\label{eq:meanlargesep}
\end{equation}
mainly because of the increasing surface radius with time.
A convenient way to demonstrate the dependence of the sound speed and
radius on the effects of stellar evolution can be illustrated in
a $(\langle\Delta\nu\rangle,\langle\delta\nu\rangle)$ diagram, depicted
in Fig.\,\ref{fig:3}. 

Additional information of structural aspects of solar-like stars is 
obtained from the seismic signatures contained in 
$\Delta\nu_{n,l}$ and $\delta\nu_{n,l}$.
Abrupt variation in the stratification of a star (relative to the scale of the 
inverse radial wavenumber of a seismic mode of oscillation), such as that 
resulting from the (smooth, albeit acoustically relatively abrupt) depression 
in the first adiabatic exponent 
$\gamma=(\partial {\ln p}/\partial{\ln\rho})_s$ caused by the ionization 
of helium, where $p$, $\rho$ and $s$ are pressure, density and specific 
entropy, or from the sharp transition from radiative to convective heat 
transport at the base of the convection zone, induces small-amplitude 
oscillatory components (with respect to frequency) in the spacing of the 
cyclic eigenfrequencies $\nu_{n,l}$ of seismic oscillation and
consequently also in $\Delta\nu_{n,l}$ and $\delta\nu_{n,l}$.
We call such abrupt variations an acoustic glitch.
One might hope that the variation of the sound speed $c$ induced by helium 
ionization might enable one to determine from the low-degree 
eigenfrequencies a measure that is directly related to, perhaps even almost 
proportional to, the helium abundance, with little contamination from other 
properties of the structure of the star.

\begin{figure}[ht]
 \centering
 \includegraphics[width=1.0\linewidth]{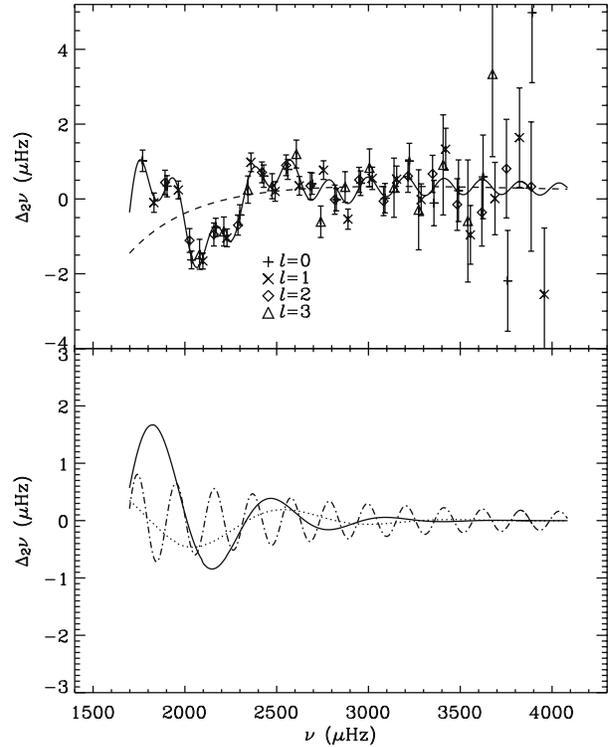}
 \caption{The symbols in the {\bf upper panel} show second differences
          $\Delta_2\nu_{n,l}$ (see equation \ref{eq:secdiff}) for low-degree 
          ($l=0,..,3$) modes obtained from simulated data for a 1\,M$_\odot$
          model of age 5.54 Gy. The simulation was based on two periods 
          of 4-months observations, separated by 1 year, with the SONG 
          network (see Grundahl et al. 2007). The solid curve is a fit to 
          $\Delta_2\nu$ based on the analysis by Houdek \& Gough (2007). The 
          dashed curve is a smooth contribution, modelled as a third-order 
          polynomial in $\nu^{-1}$, which represents near-surface 
          effects. The {\bf lower panel} displays the remaining individual 
          contributions from the acoustic glitches to $\Delta_2\nu$: the 
          dotted and solid curves show the contribution from the first and 
          second stages of helium ionization and the dot-dashed curve is 
          the contribution from the acoustic glitch at the base of 
          the convective envelope.}
 \label{fig:4}
\end{figure}

A convenient and easily evaluated measure of the oscillatory component 
produced by acoustic glitches is the second multiplet-frequency difference 
with respect to order $n$ amongst modes of like degree $l$:

\begin{equation}
\Delta_2\nu_{n,l}\equiv\nu_{n-1,l}-2\nu_{n,l}+\nu_{n+1,l}
\label{eq:secdiff}
\end{equation}
(Gough 1990). Any 
localized region of rapid variation of either the sound speed $c$ or the 
density scale height, or a spatial derivative of them, induces an 
oscillatory component in $\Delta_2\nu$ (from here on the subscripts $n,l$ 
have been dropped for simplicity) with a 
`cyclic frequency' approximately equal to twice the acoustic depth
\begin{equation}
\tau=\int_{r_{\rm glitch}}^R c^{-1}\,{\rm d}r
\end{equation}
of the glitch, and with an amplitude which depends on the amplitude of the 
glitch and which decays with $\nu$ once the inverse radial wavenumber of the 
mode becomes comparable with or less than the radial extent of the glitch.  

Various approximate formulae for the oscillatory components 
that are associated with the helium ionization 
have been suggested and used, by e.g., Basu et al. (1994, 2004),
Monteiro \& Thompson (1998, 2005) and Gough (2002), not all of which
are derived directly from explicit acoustic glitches. Gough used an
analytic function for modelling the dip in the first adiabatic exponent.
In contrast, Monteiro \& Thompson assumed a triangular form.
Basu et al. have adopted a seismic signature for helium ionization that is 
similar to that arising from a single discontinuity;
the artificial discontinuities in the sound speed and its
derivatives that this and the triangular representations possess cause 
the amplitude of the oscillatory signal to decay with frequency too gradually,
although that deficiency may not be immediately noticeable within the limited 
frequency range in which adequate asteroseismic data are or will 
imminently be available.
More recently Houdek \& Gough (2007) proposed a seismic diagnostic in which
the variation of $\gamma$ in the helium ionization zone is represented with 
a pair of Gaussian functions. This correctly results in
a decay of the amplitude of the seismic signature with oscillation
frequency that is faster than that which the triangular and the
single-discontinuity approximations imply, and also takes some account of
the two ionization states of helium. Moreover, they incorporated the
acoustic cutoff frequency into the variation of the eigenfunction phase
with acoustic depth, thereby improving the discrepancy between the
seismically inferred depths of the acoustic glitches and that of 
a corresponding stellar model. An example of the application of
Houdek \& Gough's technique is presented in Fig.\,\ref{fig:4}, which
shows the resulting fit to simulated data for a solar-like star.
\begin{figure}[t]
\centering
\includegraphics[width=0.9\linewidth]{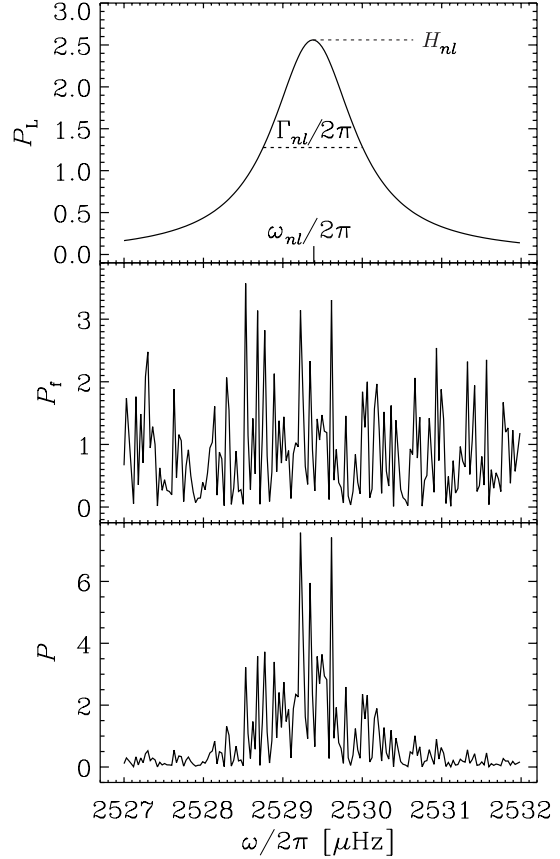}
\caption{
Power spectral density of a randomly excited, damped, harmonic oscillator.
$P_{\rm f}$ represents the spectral density of the random force
and $P$ is the product of the Lorentzian $P_{\rm L}$
and $P_{\rm f}$ (adapted from Kosovichev 1995).
}
\label{fig:5}
\end{figure}
\vspace{-12pt}
\subsection{Mode parameters}
\vspace{-8pt}
Were solar p modes to be genuinely linear and stable, their power spectrum
could be described in terms of an ensemble of intrinsically damped,
stochastically driven, simple-harmonic oscillators,
provided that the background
equilibrium state of the star is independent of time (Fig.\,\ref{fig:5});
if we assume further that mode phase fluctuations contribute negligibly to
the width of the spectral lines, the intrinsic damping rates of
the modes, $\Gamma/2$, could then be determined observationally from
measurements of the pulsation linewidths $\Gamma$.

The power (spectral density), $P$, of the surface displacement $\xi_{nl}(t)$
of a damped, stochastically driven, simple-harmonic oscillator, satisfying
\begin{equation}
I_{nl}\left[\frac{{\rm d}^2{\xi_{nl}}}{{\rm d}t^2}
+\Gamma_{nl}\frac{{\rm d}\xi_{nl}}{{\rm d}t}
+\omega_{nl}^2\xi_{nl}\right]=f(t)\,,
\label{eq:harmosc}
\end{equation}
which represents the pulsation mode of order $n$ and degree $l$, with
linewidth $\Gamma_{nl}$ and frequency $\omega_{nl}$ and mode
inertia $I_{\rm nl}$, satisfies
\begin{equation}
P\propto P_{\rm L}\,P_{\rm f}=
\frac{\Gamma_{nl}/2\pi}
{(\omega-\omega_{nl})^2+\Gamma_{nl}^2/4}\,
P_{\rm f}\,,
\label{eq:powerdensity}
\end{equation}
assuming $\Gamma_{nl}\ll\omega_{nl}$, where
$f(t)$ describes the stochastic forcing function.
Integrating equation~(\ref{eq:powerdensity}) over frequency leads to the
total mean energy in the mode
\begin{eqnarray}
I_{nl}\,V^2_{nl}\!\!\!\!&:=&\!\!\!\!
 \frac{1}{2}\omega^2_{nl}I_{nl}\langle|A_{nl}|^2\rangle
\propto\omega^2_{nl}I_{nl}\int_{-\infty}^\infty P(\omega)\,{\rm d}\omega\cr
&\propto&\frac{P_{\rm f}(\omega_{nl})}{\Gamma_{nl}}\,,
\label{eq:totalenergy}
\end{eqnarray}
where $A_{nl}$ is the displacement amplitude (angular brackets,
$\langle\rangle$, denote an expectation value) and $V_{n,l}$ is the
rms velocity of the displacement. The total mean energy of a mode is therefore
directly proportional to the rate of work (also called 
energy supply rate) of the stochastic forcing $P_{\rm f}$ at the
frequency $\omega_{nl}$ and indirectly proportional to $\Gamma_{nl}$.

Equations (\ref{eq:harmosc})--(\ref{eq:totalenergy}) are discussed in
terms of the displacement $\xi$ (from here on we omit the subscripts $n$
and $l$), but in order to have a direct relation between the observed
velocity signal $v(t)={\rm d}\xi/{\rm d}t$ and the modelled excitation
rate $P_{\rm f}$ we shall first take the Fourier transform $\tilde V(\nu)$
of $v(t)$ ($\nu=\omega/2\pi$). It follows that the total mean energy $E$
of the harmonic signal of a single pulsation mode is then given by
(Chaplin et al. 2005)
\begin{equation}
E=IV^2=I\hat\delta\int_{-\infty}^\infty\vert\tilde V(\nu)\vert^2\,{\rm d}\nu
=\frac{1}{4}I\Gamma H\,,
\label{eq:totalenergyH}
\end{equation}
in which
\begin{equation}
H:=\int_{\nu-\hat\delta/2}^{\nu+\hat\delta/2}\vert\tilde V(\nu)\vert^2\,{\rm d}\nu\,,
\label{eq:height}
\end{equation}
is the maximum power density - which corresponds to the `height' of the
resonant peak in the frequency domain (see Fig.\,\ref{fig:2}). The height $H$
is the maximum of the discrete power, i.e. the integral of power spectral
density over a frequency bin $\hat\delta=1/T_{\rm obs}$, where $T_{\rm obs}$
is the total observing time. The following expressions
\begin{equation}
V^2\,:=\,\frac{P_{\rm f}}{\Gamma I}
\,=\,\frac{1}{4}\Gamma H
\label{eq:V-H}
\end{equation}
and
\begin{equation}
H\,:=\,\frac{P_{\rm f}}{(\Gamma/2)^2I}
\label{eq:H}
\end{equation}
provide a direct relation between the observed height $H$
(in cm$^2\,$s$^{-2}$Hz$^{-1}$), the modelled
energy supply rate $P_{\rm f}$ (in erg\,s$^{-1}$), and
damping rate $\eta=\Gamma/2$.

\vspace{-10pt}
\section{Pulsation computations and amplitude ratios}
\vspace{-8pt}
\label{sec:computations}
The linearized pulsation equations for nonadiabatic radial oscillations
can be presented as (e.g. Balmforth 1992a):
\begin{eqnarray}
\frac{\partial}{\partial m}\left(\frac{\delta p}{p}\right)&=&
 \hat f\left(\frac{\delta r}{r},\frac{\delta T}{T},\frac{\delta p}{p},\frac{\delta p_{\rm t}}{p},\frac{\delta\Phi}{\Phi}\right)\,,\cr
\frac{\partial}{\partial m}\left(\frac{\delta r}{r}\right)&=&
-\frac{1}{4\pi r^3\rho}\left(3\frac{\delta r}{r}+\frac{\delta\rho}{\rho}\right)\,,\cr
\frac{\partial}{\partial m}\left(\frac{\delta L}{L}\right)&=&
-{\rm i}\omega\frac{c_pT}{L}\left(\frac{\delta T}{T}-\nabla_{\rm ad}\frac{\delta p}{p}\right)\,,
\label{eq:pulsation}
\end{eqnarray}
where $\delta$ is the Lagrangian perturbation operator, and for simplicity the
right hand side of the perturbed momentum equation is formally expressed by the
function $\hat f$ (the full set of equations can be found in, e.g.,
Balmforth 1992a).
Equations~(\ref{eq:pulsation}) are solved subject to boundary conditions to
obtain the eigenfunctions and the complex angular eigenfrequency
$\omega=\omega_{\rm r}+{\rm i}\eta$, where $\omega_{\rm r}$ is the (real)
pulsation frequency and $\eta=\Gamma/2$ is the damping rate in (s$^{-1}$).
The turbulent flux perturbations of heat and momentum, $\delta L_{\rm c}$ and
$\delta p_{\rm t}$, and the fluctuating anisotropy factor $\delta\Phi$ are
obtained from the nonlocal, time-dependent convection formulation
by Gough (1977a,b).

From the linearized nonadiabatic pulsation equations 
(\ref{eq:pulsation}) theoretical intensity-velocity amplitude ratios
\begin{equation}
\frac{\Delta L_{\rm s}}{\Delta V}:=
                        \frac{\delta L/L}{\omega_{\rm r} r\;\delta r/r}
\label{eq:amprat}
\end{equation}
can be compared with observations, without the need of a specific
excitation model and all its uncertainties in describing the turbulent 
spectra.

It is, however, important to realize that various instruments observe in
different absorption lines and consequently at different heights in the
atmosphere. This property has to be taken into account not only when comparing
observations between various instruments 
(e.g. Christensen-Dalsgaard \& Gough 1982), but also when comparing theoretical
amplitude estimates with observations (Houdek et al. 1995). Table~1
lists some of the relevant properties of various instruments.

\begin{table}
\caption{Absorption lines and their wavelengths $\lambda$ of various
helioseismic instruments. Also listed are the optical depths $\tau_{5000}$
at 5000\,\AA\ and the corresponding approximate heights above the photosphere
($h=0$ at $T=T_{\rm eff}$) at which the lines are formed.
\bigskip
}
\renewcommand{\arraystretch}{1.2}
\bigskip
\begin{tabular}[ht]{lcclc}
Instrument$\!\!\!$&$\!\!\!\!\!\!\!$line$\!\!\!\!\!\!$&$\lambda\,$(\AA)&$\tau_{5000}$&$\!\!\!\!\!\!$height\,(km)$\!\!\!\!$\\
\noalign{\medskip}
\hline
\noalign{\medskip}
BBSO &Ca      &6439&0.05            &$\sim129$\tablenote{from Libbrecht (1988)}\\
BiSON&K       &7699&0.013           &$\sim250$\tablenote{from Christen-Dalsgaard \& Gough (1982)}\\
MDI  &Ni\,I   &6708&9$\times10^{-3}$&$\sim300$\tablenote{from Toutain et al. (1997, but see also Baudin et al. 2005)}\\
GOLF &Na D1/D2&5690&5$\times10^{-4}$&$\ \ \,\,\,\sim500^{**}$\\
\end{tabular}
\end{table}

\begin{figure}[t]
\centering
\includegraphics[width=0.96\linewidth]{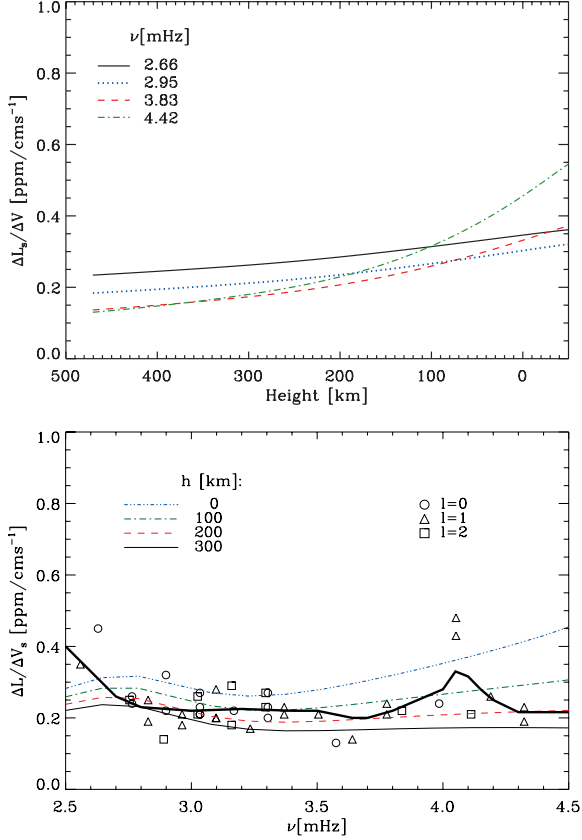}
\caption{
{\bf Top}:
Calculated amplitude ratios (see equation (\ref{eq:amprat})) as a function
of height in a solar model for modes with different frequency values.
{\bf Bottom}:
Theoretical amplitude ratios (surface luminosity perturbation over velocity)
for a solar model compared with observations by Schrijver et\,al. (1991).
Computed results are depicted at different heights above the photosphere
($h$=0\,km at $T=T_{\rm eff}$) The thick, solid curve indicates a
running-mean average of the data (from Houdek et~al. 1995).
}
\label{fig:amprat}
\end{figure}

\begin{figure}[t]
\centering
\includegraphics[width=0.96\linewidth]{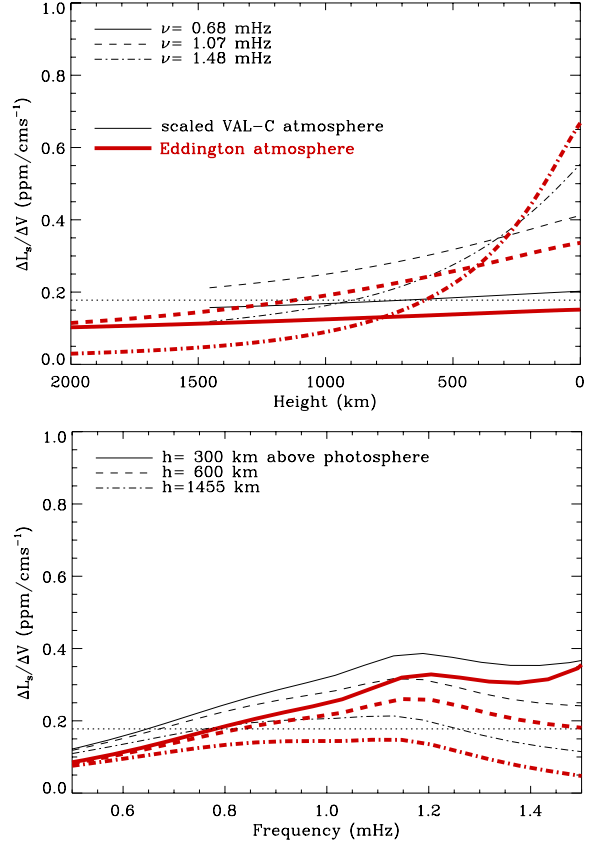}
\caption{Calculated amplitude ratios (see equation (\ref{eq:amprat})) 
for a model of Procyon A are compared with observations
by Arentoft et al. (2008; horizontal dotted line). Theoretical results are
shown for a scaled VAL-C atmosphere (black, thin curves) and for an 
Eddington atmosphere (red, thick curves).
{\bf Top}: 
The theoretical amplitude ratios are shown as a function of height and
for three different pulsation modes. The frequencies of the three 
pulsation modes are indicated.
{\bf Bottom}:
The theoretical amplitude ratios are shown as a function of frequency
at three different heights in the stellar atmosphere. The heights 
above the photosphere $h=0\,$km are indicated. }
\label{fig:amprat_procyon}
\end{figure}

In the top panel of Fig.~\ref{fig:amprat} the theoretical amplitude ratios
(equation~(\ref{eq:amprat})) of a solar model are plotted as a function of
height for several radial pulsation modes. The mode energy density
(which is proportional to $r\rho^{1/2}\delta r$) increases rather slowly with
height; the density $\rho$, however, decreases very rapidly and consequently
the displacement eigenfunction $\delta r$ increases with height. This leads
to the results shown in the upper panel of Fig.~\ref{fig:amprat} where the
decrease in the amplitude ratios with height is particularly pronounced for
high-order modes for which the eigenfunctions vary rapidly in the evanescent
outer layers of the atmosphere. It is for that reason why solar velocity
amplitudes from, e.g., the GOLF instrument have larger values than the
measurements from the BiSON instrument (by about 25\%, Kjeldsen et~al. 2005).\\
The lower panel of Fig.~\ref{fig:amprat} compares the estimated solar amplitude
ratios (curves) with observed ratios (symbols) as a function of frequency. The
model results are depicted for velocity amplitudes computed at different
atmospheric levels. The observations are obtained from accurate irradiance
measurements from the IPHIR instrument of the PHOBOS 2 spacecraft with
contemporaneous low-degree velocity data from the BiSON instrument at
Tenerife (Schrijver et al. 1991). The thick solid curve represents a
running-mean average, with a width of 300$\,\mu$Hz, of the observational data.
The theoretical ratios for $h=200\,$km (dashed curve) show reasonable
agreement with the observations.

In Fig.~\ref{fig:amprat_procyon} model results for the F5 star Procyon A are 
compared with observations (horizontal dotted line) by Arentoft et al. (2008). 
Theoretical results are shown for two stellar atmospheres: a VAL-C 
(Vernazza et al. 1981) atmosphere scaled with the model's effective
temperature $T_{\rm eff}$ (thin curves), and for an Eddington atmosphere
(thick curves). For both stellar atmospheres the agreement with the 
observations is less satisfactory than in the solar case, indicating 
that we may not represent correctly the shape of the pulsation eigenfunctions.
Consequently there is need for adopting more realistically computed 
atmospheres in the equilibrium models, particularly for stars with much higher
surface temperatures than the Sun. It should, however, be mentioned that the
current photometric observations are still uncertain.

\begin{figure}[t]
\centering
\includegraphics[width=1.00\textwidth]{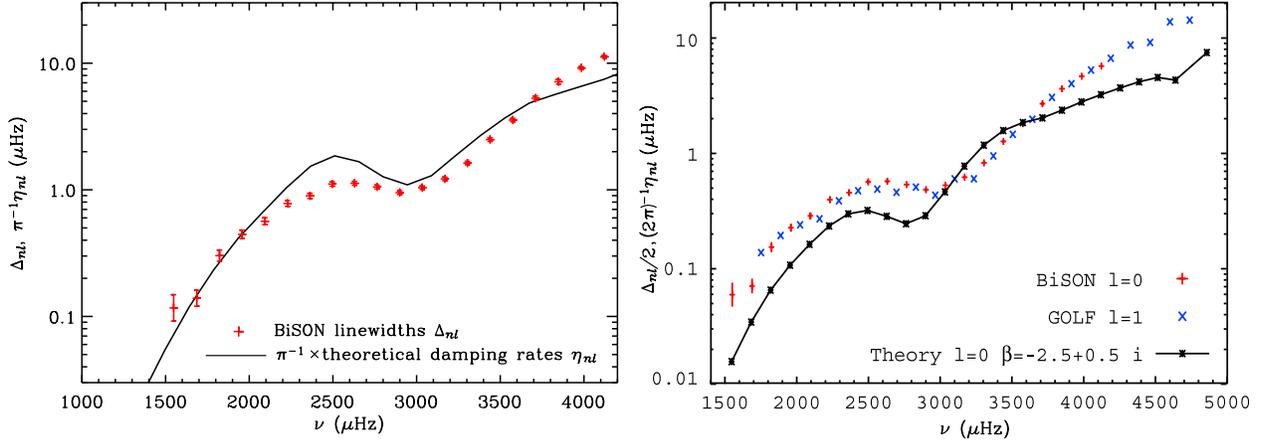}
\caption{
The symbols in the left-hand panel are the measured linewidths 
$\Delta_{nl}=\Gamma_{nl}/2\pi$ (we denote the FWHM in unit of cyclic
frequency by $\Delta_{nl}$) of solar low-degree p modes obtained from a 3456-d
data set collected by BiSON between 1991 and 2000 (Chaplin et al. 2005).
The data are compared with the theoretical damping rates $\pi^{-1}\eta_{nl}$ 
(connected by the solid curve) obtained from the model 
computations discussed in the previous section (from Chaplin et al. 2005). 
In the right-hand panel theoretical results of $(2\pi)^{-1}\eta_{nl}$ 
(solid curve) by Dupret et al. (2004) are compared with observations 
of $\Delta_{nl}/2$ (symbols).
}
\label{fig:damping}
\end{figure}

\vspace{-10pt}
\section{Damping rates}
\vspace{-8pt}
\label{sec:solar-damping}
Damping of stellar oscillations arises basically from two sources: processes
influencing the momentum balance, and processes influencing the thermal energy
equation. Each of these contributions can be divided further according to their
physical origin, which was discussed in detail by Houdek et al. (1999).

Important processes that influence the thermal energy balance are
nonadiabatic processes attributed to the modulation of the convective heat
flux by the pulsation. This contribution is related to the way that convection
modulates large-scale temperature perturbations induced by the pulsations
which, together with the conventional $\kappa$-mechanism, influences
pulsational stability.

Current models suggest that an important contribution that influences
the momentum balance is the exchange of energy between the pulsation and
the turbulent velocity field through dynamical effects of the fluctuating
Reynolds stress. In fact, it is the modulation of the turbulent fluxes by
the pulsations that seems to be the predominant mechanism responsible for
the driving and damping of solar-type acoustic modes.
It was first reported by Gough (1980) that the dynamical effects arising
from the turbulent momentum flux perturbation $\delta p_{\rm t}$
contribute significantly to the damping $\Gamma$.
Detailed analyses (Balmforth 1992a) reveal how damping is controlled largely
by the phase difference between the momentum perturbation and the density
perturbation. Therefore, turbulent pressure fluctuations must not be neglected
in stability analyses of solar-type p modes.

A comparison between the latest linewidth measurements (full-width at
half-maximum) $\Delta_{nl}=\Gamma_{nl}/2\pi$ and theoretical damping
rates is given in Fig.~\ref{fig:damping}. The observational time series
from BiSON (Chaplin et al. 2005) was obtained from a 3456-d data set and
the linewidths of the temporal power spectrum extend over many
frequency bins $\hat\delta=1/2T_{\rm obs}$. In that case the linewidth in
units of cyclic frequency is related to the damping rate according to
\begin{equation}
\Delta_{nl}=\pi^{-1}\eta_{nl}\,.
\end{equation}

Recently Dupret et al. (2004) performed similar
stability computations for the Sun using the time-dependent mixing-length
formulation by Gabriel et al. (1975, 1998) and Grigahc\`ene et al. (2005),
which is based on the formulation by Unno (1967). Their results
are illustrated in the
right panel of Fig.~\ref{fig:damping}, which also shows the characteristic
plateau near 2.8\,mHz. It is, however, interesting to note that their findings
suggest the fluctuating convective heat flux to be the main contribution
to mode damping, whereas for the model results shown in the left panel of
Fig.~\ref{fig:damping}, which are based on Gough's (1977a,b) convection
formulation, it is predominantly the fluctuating Reynolds stress that
makes all modes stable.

\begin{figure}[t]
\centering
\includegraphics[width=1.00\linewidth]{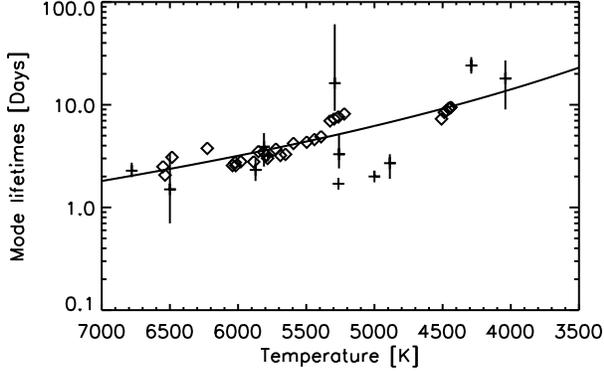}
\caption{Average mode lifetimes $\langle\tau\rangle$ (averaged over 
the five most prominent p modes) in low-mass stars. The diamond symbols 
are the results
from the stellar equilibrium and pulsation calculations and the cross
symbols are observations of 12 main-sequence, sub-giant and red-giant 
stars. The solid curve represents the power law $T^{-4}_{\rm eff}$, where
$T_{\rm eff}$ is the surface temperature of the models that were used in 
the stability computations (from Chaplin et al. 2009).
} 
\label{fig:9}
\end{figure}

Houdek et al. (1999) computed damping rates $\eta$ of solar-like oscillations 
in about 160 stars with masses between 0.9\,M$_\odot$ and 2.0\,M$_\odot$ in 
the vicinity of the main sequence, and for various metallicities and 
convection parameters. Recently Chaplin et al. (2009)
conducted a similar theoretical study of estimating mode lifetimes
$\tau=\eta^{-1}$ in low-mass stars and compared the
theoretical estimates with the latest linewidth measurements in twelve 
solar-type variables. They found that the mean mode lifetimes of the five 
most prominent solar-like p modes $\langle\tau\rangle$ scale like
\begin{equation}
\langle\tau\rangle\propto T^{-4}_{\rm eff}\,.
\label{eq:meantau}
\end{equation}
Their results are depicted in Fig.\,\ref{fig:9}, where the diamond symbols
are the theoretical estimates of $\langle\tau\rangle$ obtained from a grid 
of models computed in the manner of Chaplin et al. (2005) and the crosses 
with error bars are the linewidth measurements of twelve main-sequence, 
sub-giant and red-giant stars.

\vspace{-10pt}
\section{Stochastic excitation}
\vspace{-8pt}
\label{sec:excitation}
Because of the lack of a complete model for convection, the mixing-length
formalism still represents the main method for computing the turbulent
fluxes in the convectively unstable layers in a star. One of the assumptions
in the mixing-length formulation is the Boussinesq approximation, which
results in neglecting the acoustic wave generation by assuming the fluid to
be incompressible. Consequently a separate model is needed to estimate the
rate of the acoustic noise (energy supply rate) generated by the turbulence.
The excitation process can be regarded as multipole acoustic radiation
(Lighthill 1952).
Acoustic radiation by turbulent multipole sources in the context of stellar
aerodynamics has been  considered by Unno \& Kato (1962), Moore \&
Spiegel (1964), Unno (1964), Stein (1967),  Goldreich \& Keeley (1977),
Bohn (1984), Osaki (1990), Goldreich \& Kumar (1990), Balmforth (1992b),
Goldreich, Murray \& Kumar (1994), Musielak et~al. (1994),
Samadi \& Goupil (2001) and Chaplin et~al. (2005).

The mean amplitude $A$ of a mode is determined by a 
balance by the
energy supply rate $P_{\rm f}$ from the turbulent velocity field
and the thermal and mechanical dissipation rate characterized by
the damping coefficient $\eta$ [see equation (\ref{eq:totalenergy})].
The procedure that we adopt to estimate $A$ is that 
of Chaplin et~al. (2005),
whose prescription follows that of Balmforth (1992b).

We represent the linearized pulsation dynamics 
by the simplified equation
\begin{equation}
\rho\left(\frac{\partial^2\bm\xi}{\partial t^2}
         +2\eta\frac{\partial\bm\xi}{\partial t}
         +\mathcal{L}\bm\xi\right)=
\bm{F}(\bu)+\bm{G}(s^{\prime})
\label{eq:excitation}
\end{equation}
for the displacement $\bm\xi(\bm r,t)$, which is now
also a function of radius $\bm r$, of a forced oscillation
corresponding to a single radial mode satisfying the homogeneous equation
\begin{equation}
{\cal L}\bm\xi(\bm r)=
\omega^2\bm\xi(\bm r)
\end{equation}
in which $\omega$ (and $\bm\xi$) are real and ${\cal L}$ is
a linear spatial operator. The (inhomogeneous) fluctuating terms on
the right-hand-side of equation~(\ref{eq:excitation}) arise from the
fluctuating Reynolds stresses
\begin{equation}
\bm{F}(\bu)=\nabla\cdot(\rho\bu\bu-
\langle\rho\bu\bu\rangle)
\label{eq:F}
\end{equation}
and from the fluctuating gas pressure (due to the fluctuating buoyancy force),
represented by $\bm{G}(s^{\prime})$, where $s^{\prime}$
is the Eulerian entropy fluctuation (Bohn 1984; Osaki 1990;
Goldreich \& Kumar 1990; Balmforth 1992b;
Goldreich, Murray \& Kumar 1994; Samadi \& Goupil 2001).
The latest numerical simulations by
Stein et al. (2004) suggest that both forcing terms in
equation~(\ref{eq:excitation}) contribute to the energy supply rate $P_{\rm f}$
by about the same amount, a result that
was also reported by Samadi et al. (2003) using the turbulent velocity
field and anisotropy factors from numerical simulations
(Stein \& Nordlund 2001).
In this paper we consider only the term of the fluctuating Reynolds stresses
and because we use only radial modes, only the vertical
component $F_3$ of $\bm{F}$ is important,
\begin{equation}
F_3(u_3)\simeq
\frac{\partial}{\partial r}(\rho u^2_3-\langle\rho u^2_3\rangle)\,.
\label{eq:F3}
\end{equation}
If we define the vertical component of the velocity correlation as
$R_{33}=\langle u_3u_3\rangle$, its Fourier transform $\widehat R_{33}$ can
be expressed in the Boussinesq-quasi-normal approximation
(e.g. Batchelor 1953) as a function of the turbulent energy spectrum function
$E(k,\omega)$:
\begin{equation}
\widehat R_{33}=\frac{\Psi E(k,\omega)}{12\pi k^2}\,,
\end{equation}
where $k$ is a wavenumber and $\Psi$ is an anisotropy parameters given by
\begin{equation}
\Psi=\left[{2\Phi}/{3(\Phi-1)}\right]^{1/2}\,,
\end{equation}
which is unity for isotropic turbulence (Chaplin et~al. 2005). This factor
was neglected in previously published excitation models but it has to be
included in a consistent computation of the acoustic energy supply rate.
Following Stein (1967) we factorize the energy spectrum function into
$E(k,\omega)=\tilde E(k)\Omega(\omega; \tau_k)$, where $\tau_k=\lambda/ku_k$
is the correlation time-scale of eddies of size $\pi/k$ and velocity $u_k$;
the correlation factor $\lambda$ is of order unity and accounts for
uncertainties in defining $\tau_k$.

\begin{figure}[t]
\centering
\includegraphics[width=1.00\linewidth]{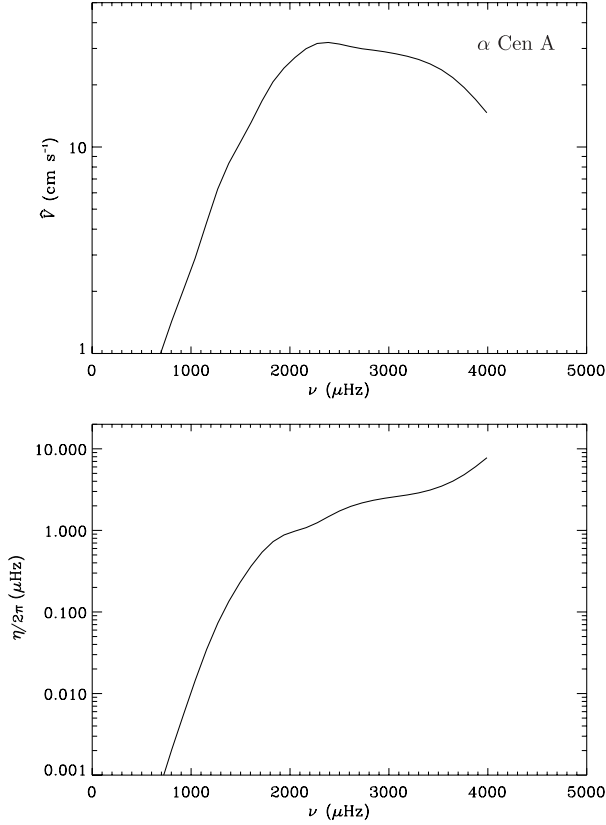}
\caption{
{\bf Top}:
Predicted apparent velocity amplitudes (defined to be $\sqrt2$ times the 
rms value) for a model of $\alpha$~Cen~A, computed according to 
equation~(\ref{eq:V-H}).  
{\bf Bottom}:
Linear damping rates for a model of $\alpha$~Cen~A, obtained by solving 
the fully nonadiabatic pulsation equations~(\ref{eq:pulsation}).
} 
\label{fig:aCenA}
\end{figure} 
\begin{figure}[t]
\centering
\includegraphics[width=1.00\linewidth]{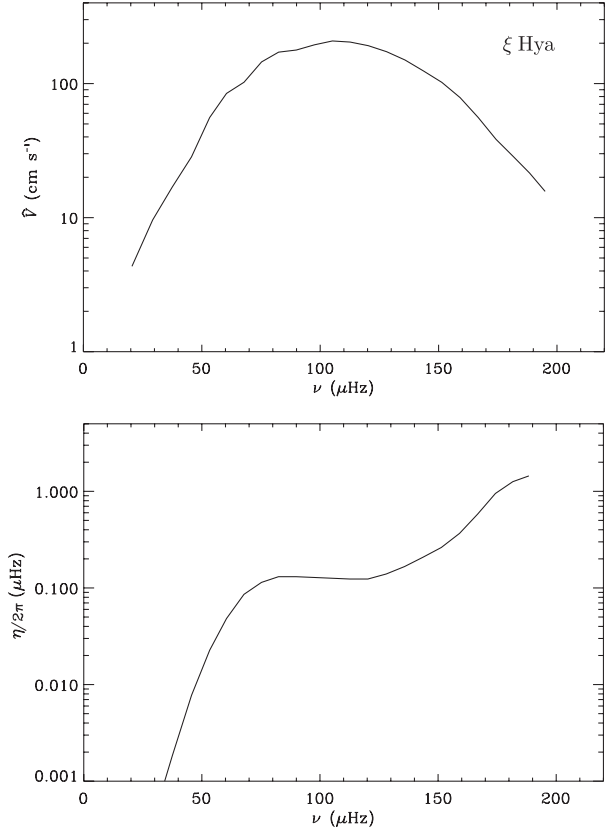}
\caption{
{\bf Top}: 
Predicted apparent velocity amplitudes (defined to be $\sqrt2$ times the 
rms value) for a model of $\xi$~Hydrae, computed according to 
equation~(\ref{eq:V-H}).
{\bf Bottom}:
Linear damping rates for a model of $\xi$~Hydrae, obtained by solving the 
fully nonadiabatic pulsation equations~(\ref{eq:pulsation}) 
(adapted from Houdek \& Gough 2002).
}
\label{fig:xiHya}
\end{figure} 

The energy supply rate is then given by (see Chaplin et~al. 2005 for details)
\begin{equation}
P_{\rm f}=
    \frac{\pi}{9I}
    \int_0^R \ell^3
    \left(\Phi\Psi rp_{\rm t}\frac{\partial{\xi}_{r}}{\partial r}\right)^2
    {\cal S}(r;\omega)\,{\rm d}r\,,
\label{eq:excitation-rate}
\end{equation}
with
\begin{equation}
{\cal S}(r;\omega)=\int_0^\infty \kappa^{-2}\tilde E^2(\kappa)
                     \tilde\Omega(\tau_k;\omega)\,{\rm d}\kappa\,,
\label{eq:function-S}
\end{equation}
where $\kappa=k\ell/\pi$, $\ell$ is the mixing length, $R$ is surface radius,
and $\xi_r$ is the normalized radial part of $\bm{\xi}$.
The spectral function ${\cal S}$ accounts for contributions to $P_{\rm f}$
from the small-scale turbulence and includes the normalized spatial
turbulent energy spectrum $\tilde E(k)$ and the
frequency-dependent factor $\Omega(\tau_k;\omega)$. For $\tilde E(k)$
it has been common to adopt either the Kolmogorov (Kolmogorov 1941)
or the Spiegel spectrum (Spiegel 1962). The frequency-dependent factor
$\Omega(\tau_k;\omega)$ is still modelled in a very rudimentary way
and we adopt two forms:\\
-- the Gaussian factor (Stein 1967)\,,
\begin{equation}
\Omega_{\rm G}(\omega;\tau_k)=
    \frac{\tau_k}{\sqrt{2\pi}}\,{\rm e}^{-(\omega\tau_k/\sqrt2)^2}\,;
\label{eq:Gaussian}
\end{equation}
-- the Lorentzian factor
   (Gough 1977b; Samadi et al. 2003; Chaplin et al. 2005)\,,
\begin{equation}
\Omega_{\rm L}(\omega;\tau_k)=
 \frac{\tau_k}{\pi\sqrt{2\ln2}}\,\frac{1}{1+(\omega\tau_k/\sqrt{2\ln2})^2}\,.
\label{eq:Lorentzian}
\end{equation}
The Lorentzian frequency factor is a result predicted for the largest,
most-energetic eddies by the time-dependent mixing-length formulation
of Gough~(1977b).  Recently, Samadi et al. (2003) reported that
Stein \& Nordlund's hydrodynamical simulations also suggest a Lorentzian
frequency factor, which decays more slowly
with depth $z$ and frequency $\omega$ than the Gaussian factor. Consequently
a substantial fraction to the integrand of
equation~(\ref{eq:excitation-rate}) arises from eddies situated in the deeper
layers of the Sun, resulting in a larger acoustic excitation
rate $P_{\rm f}$.

\begin{figure}[t]
\centering
\includegraphics[width=1.00\linewidth]{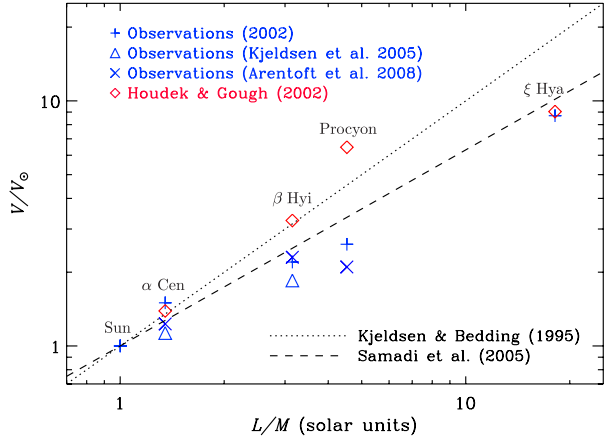}
\caption{
Predicted apparent velocity amplitudes (defined to be $\sqrt2$ times the rms 
value) as function of light-to-mass ratio for stochastically excited 
oscillations
in other stars. Observations from several authors are plotted by the
plus and triangle symbols. The theoretical estimates by Houdek \& Gough (2002)
are plotted as diamond symbols. The scaling law of Kjeldsen \& Bedding (1995) 
is illustrated by the dotted line and results reported by 
Samadi et~al. (2005) are indicated by the dashed line.
} 
\label{fig:veloamp}
\end{figure} 

\vspace{-10pt}
\section{Oscillation amplitudes}
\vspace{-8pt}
\label{sec:amplitudes}
Model predictions of the mode height $H$ were computed according to 
equation~(\ref{eq:H}).
As in the solar case damping rates for other stars are obtained from
solving the eigenvalue problem~(\ref{eq:pulsation}) and
the energy supply rates are calculated from 
expression~(\ref{eq:excitation-rate}).
With these estimates for $\eta$ and $P_{\rm f}$ Houdek \& Gough (2002)
predicted velocity amplitudes for several stars, using 
equation~(\ref{eq:V-H}). Results for stochastically
excited oscillation amplitudes and linear damping rates in the solar-like
star $\alpha$\,Cen\,A and in the sub-giant $\xi$\,Hydrae are illustrated
in Figs\,\ref{fig:aCenA} and \ref{fig:xiHya}.
Kjeldsen\,et\,al.\,(2005) reported
mode lifetimes for $\alpha$\,Cen\,A of about 2.1 days (but see also
Fletcher et al. 2006), which are in reasonable agreement with the 
theoretical estimates of about $1.7$ days for the most prominent modes 
(the mode lifetime $\tau\!=\!\eta^{-1}$; see lower panel of 
Fig.~\ref{fig:aCenA}).
For $\xi$~Hydrae, however, the theoretical mode lifetime of the most 
prominent modes is about 18 days which is in stark contrast 
to the measured value of about 2--3 days by Stello et~al. (2004, 2006), 
yet the estimated velocity amplitudes for $\xi\,$ Hydrae are in almost 
perfect agreement with the observations by Frandsen et~al. (2002).

A comparison between predicted and observed velocity amplitudes in several
solar-type pulsators is
illustrated in Fig.~\ref{fig:veloamp}. The dotted line is the scaling
law by Kjeldsen \& Bedding (1995), which is based on the computations
by Christensen-Dalsgaard \& Frandsen (1983), and the dashed line is the 
scaling relation reported by Samadi et~al. (2005) using the convective 
velocity profiles from 3D numerical simulations (Stein \& Nordlund 2001), a
Lorentzian frequency factor in equation~(\ref{eq:function-S}), and the
theoretical damping rates from Houdek et~al. (1999). 
For the cooler stars the theoretical results (scaling laws) are in 
reasonable agreement with the observations, whereas for hotter stars, such as
for Procyon A, the theoretical velocity amplitudes are overestimated by both 
the scaling laws and the stochastic excitation models.

Recently, Chaplin et al. (2009) suggested a new scaling law for the intensity
amplitudes by combining their finding of the mean mode lifetime,
$\langle\tau\rangle\propto T^{-4}_{\rm eff}$ 
[cf. equation\,(\ref{eq:meantau})], 
with Kjeldsen \& Bedding's (1995) scaling law for the intensity
amplitudes inferred from a narrow-band observation of wavelength $\lambda$,
i.e. $(\delta L/L)_\lambda\propto(L/M)/T^2_{\rm eff}$, leading to
\begin{equation}
H\propto g^{-2}
\label{eq:newHscaling}
\end{equation}
for the maximum mode height. Since the surface gravity changes fairly slowly
along the main sequence, the new scaling relation (\ref{eq:newHscaling}) for 
$H$, which assumes narrow-band intensity observations, suggests that stars 
notably cooler than the Sun might have mode heights that are comparable
to those solar-like pulsators that are hotter than the Sun.
Moreover, Hekker et al. (2009) reported that the intensity heights
in about 780 red giant stars, observed in broadband photometry with the CoRoT
satellite, follow the scaling law $H\propto g^{-2.2}$, a result that supports 
the theoretical finding (\ref{eq:newHscaling}) by Chaplin et al. 2009. 

\begin{figure}[t]
\centering
\includegraphics[width=0.95\textwidth]{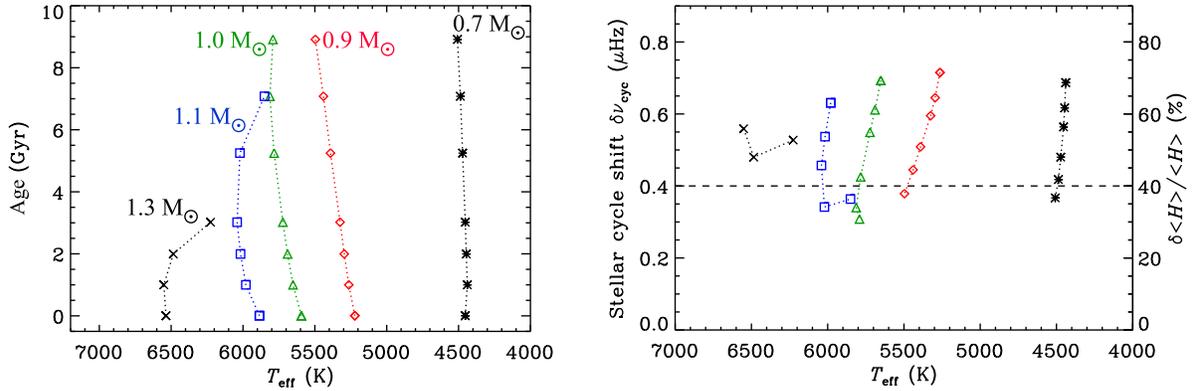}
\caption{ {\bf Left}: Age -- effective temperature plot for a sequence of
stellar models, obtained from the Padova isochrones, for which 
the stellar-cycle induced frequency shifts and mode height variations were
predicted.
{\bf Right}: Predicted stellar-cycle frequency shifts (left ordinate) and
relative height variations (right ordinate) for a sequence of stellar models
with masses between 0.7\,M$_\odot$ and 1.3\,M\,$_\odot$ 
(from Chaplin et al. 2008).} 
\label{fig:stellar-cycle}
\end{figure} 
\vspace{-10pt}
\section{Stellar-cycle effects}
\vspace{-8pt}
From helioseismic data, such as those provided by the BiSON, we 
have learnt that not only the 
oscillation frequencies change with time over the 11-year solar cycle
but also the height $H$ and width $\Gamma$. In the solar case the absolute 
fractional change from solar activity minimum to maximum in 
$\langle H\rangle$ (angular brackets $\langle\rangle$ denote 
an average over the five most prominent p modes) is about 40\%. That
in $\langle\Gamma\rangle$ is about 20\% (Chaplin et al. 2000).
We therefore expect also in solar-like oscillators not only the mode 
frequencies but also the mode heights and linewidths to show stellar-cycle 
variations. 

Chaplin et al. (2007, 2008) estimated  stellar-cycle frequency
shifts and mode height variations for a grid of 31 models with masses
between 0.7\,M$_\odot$ and 1.3\,M\,$_\odot$ and stellar ages, $t_\star$,
in the range from the ZAMS to 9\,Gy, using the Padova isochrones 
(e.g., Bonatto et al. 2004) to specify mass, radius, 
effective temperature, $T_{\rm eff}$, and luminosity. The evolutionary tracks 
of these models are presented in the left panel of 
Fig.\,\ref{fig:stellar-cycle} in terms of a ($t_\star$, $T_{\rm eff}$) plot.
To predict the stellar-cycle changes Chaplin et al. used 
the Ca II H\&K index for surface activity in stars. This index
is usually expressed as $R^\prime_{\rm HK}$, the average fraction of the 
star's total luminosity that is emitted in the H\&K line cores. The authors
then used the data of 22 main-sequence stars collected by the
Mount Wilson Ca II H\&K programme (e.g., Saar \& Brandenburg 2002) 
from which the $R^\prime_{\rm HK}$ cycle amplitude 
values, $\Delta R^\prime_{\rm HK}$, were determined. The 
$\Delta R^\prime_{\rm HK}$ values were then simply scaled against the 
0.4\,$\mu$Hz frequency shift seen for the most prominent low-$l$ modes in the
Sun, assuming that the frequency shifts $\delta\nu_{\rm cyc}$ scale 
approximately linearly with
$\Delta R^\prime_{\rm HK}$. In order to estimate the frequency shifts of
the 31 grid models, it was necessary to calculate first the $R^\prime_{\rm HK}$
values for the 31 models. This was done according to the procedure by
Noyes (1983) and Noyes et al. (1984). The $R^\prime_{\rm HK}$ values so 
obtained were then used to estimate the corresponding cycle frequency shifts by 
interpolating linearly in the ($\delta\nu_{\rm cyc}$, $R^\prime_{\rm HK}$) 
table of the 22 main-sequence stars observed by the Mount Wilson Ca II H\&K programme (for a detailed discussion see Chaplin et al. 2007). The outcome
of this procedure for estimating the stellar-cycle frequency shifts of the
most prominent p modes in the stellar models is depicted in the
right panel (left ordinate) of Fig.\,{\ref{fig:stellar-cycle}}. Also shown 
in this figure are the predicted fractional changes in $\langle H\rangle$ 
(right ordinate), obtained in a similar way as the frequency shifts and
assuming a fractional change of $\langle H\rangle$ of 40\% for the Sun.
These results suggest that, in the considered age range of 1 -- 9\,Gy,
the variation of p-mode frequencies and heights over the stellar cycle
can be up to 1.5 -- 2 times larger than in the Sun, and that these
variations depend predominantly on stellar age and less on $T_{\rm eff}$
or stellar mass.


\vspace{-10pt}
\begin{theacknowledgments}
\vspace{-8pt}
I am grateful to Douglas Gough for helpful discussions. Support
by the Austrian Science Fund (FWF project P21205) is gratefully 
acknowledged. 
\end{theacknowledgments}


\end{document}